\documentclass[12pts,onecolumn,journal]{IEEEtran}
\IEEEoverridecommandlockouts
\usepackage{cite}
\usepackage{amsmath,amssymb,amsfonts}
\usepackage{algorithmic}
\usepackage{graphicx}
\usepackage{textcomp}
\usepackage{xcolor}

\def\BibTeX{{\rm B\kern-.05em{\sc i\kern-.025em b}\kern-.08em
    T\kern-.1667em\lower.7ex\hbox{E}\kern-.125emX}}
\begin{document}
\renewcommand{\baselinestretch}{2.0} 
\title{Double Threshold based Optimal Device Selection Scheme for D2D or Sidelink Network\thanks{Part of this work was presented in \em{11th Annual Computing and Communication Workshop and Conference (IEEE-CCWC)}, 2021.}
}

\author{\IEEEauthorblockN{1\textsuperscript{st} Shamganth Kumarapandian}
\IEEEauthorblockA{\textit{Department of Engineering} \\
\textit{University of Technology and Applied Sciences, Ibra, Oman}\\
 shamkanth@ict.edu.om }\\
\and
\IEEEauthorblockN{2\textsuperscript{nd} Qasim Zeeshan Ahmed}
\IEEEauthorblockA{\textit{School of Computing and Engineering} \\
\textit{University of Huddersfield, Huddersfield, UK}\\
q.ahmed@hud.ac.uk}\\
\and
\IEEEauthorblockN{3\textsuperscript{rd} Faheem A. Khan}
\IEEEauthorblockA{\textit{School of Computing and Engineering} \\
\textit{University of Huddersfield, Huddersfield, UK}\\
f.khan@hud.ac.uk}
}

\maketitle

\begin{abstract}
Device-to-device (D2D) or Sidelink aided communication is regarded as one of the most promising technologies to improve the spectral efficiency of the $5$G and beyond communication system. However, two main challenges exist: 1) the selection of the optimal number of devices for improving the spectral efficiency, and 2) improving the physical layer security of such a communication system. The optimal device improves the secrecy capacity, and the selection of optimal devices enhances the physical layer security. Therefore, in this paper, we propose a double threshold-based optimal device selection (ODS) scheme for a cooperative wireless network with amplify and forward (AF) protocol in the presence and absence of an eavesdropper to enhance the physical layer security for the D2D network. The proposed scheme is analyzed with different distance cases, device scenarios, and modulation schemes. The bit error rate (BER) performance analysis concludes that a performance gain of more than $4$~dB is achieved at the BER of $0.004$ for the proposed double threshold-based ODS scheme compared to the existing ODS scheme at a high signal to noise ratio (SNR). Furthermore, the proposed scheme enhances the physical layer security.   
\end{abstract}

\section{Introduction}
D2D communication is one of the most promising techniques in $5$G and beyond cellular network technologies~\cite{Asadi-2014, Pan-2017,Pan-2017a, Pan-2017b,Pizzi-2021, Ahmed-2021, Husbands-2017}. This technology will result in achieving ultra-low latency, high data rates, and ultra-high reliability~\cite{Boccardi-2014}. However, these benefits come at the cost of a security breach due to the presence of anonymous nearby devices that are involved in transferring the information from the source to the end device~\cite{Dong-2010}. The other main issue with the D2D networks is to select those nearby devices that can maximize the capacity to transfer the source information to the end device~\cite{Jameel-2018}. Device selection plays an important role to enhance the performance of cooperative wireless networks~\cite{Ahmed-2015, Ahmed-2012, Ahmed-2014a}. Ineffective device selection results in the degradation of the overall performance of the network~\cite{Jameel-2019, Zou-2013}. Therefore, in this paper, new techniques that are critical to improving the physical layer security as well as device selection for the D2D or Sidelink networks are proposed. Furthermore, these techniques can easily scale when the number of devices is increased or decreased and adapt to different modulation schemes. Such techniques will also complement the existing application-layer cryptography-based security techniques and device selection for content preference and shared willingness.

Eavesdropping attack in D2D communication is one of the possible security attacks in D2D network  [4] and this attack is considered in this paper. The eavesdropper node considered in this work is passive eavesdropper in which the eavesdropper node located near to the relay node involved in accessing the source information. The chances of eavesdropper to access the information transferred from the relays to the destination is simple if the capacity of the relay to destination channel is less than relay to eavesdropper channel [5].

However, as the cellular network can be classified as network- or device-centric.  Network-centric involves the network infrastructure for all the communication between mobile users. On the other hand, in device-centric, communication is managed by the proximate device itself [2]. In cellular networks, D2D communication provides network offloading by direct communication between mobile subscribers without using core network elements [3]. Direct D2D communication has partial or full control from the base station. D2D communication is broadly catagorized as In-band D2D and Out-band D2D. In-band D2D uses the licensed spectrum, and the Out-band D2D uses an unlicensed spectrum for communication among multiple mobile devices in the network[4]. 

Secured communication is essential for both infrastructure-oriented and infrastructure-less networks. Communication security in out-band device-to-device (D2D) network is critical because of its infrastructure-less network type. It is also due to the participation of anonymous relays that are involved in the transfer of the source signal to the destination node.

\subsection{Related Works}
To minimize the outage probability of a cooperative wireless network, a scheme employing AF protocol was proposed in~\cite{Park-2013} where a single relay and an active eavesdropper with cooperative jamming were considered. The AF-based optimal relay selection was proposed in~\cite{Zou-2013} to improve the security in presence of a passive eavesdropper and multiple relays or devices. The intercept probability performance was studied and it was shown that the proposed system performed better than the traditional relay selection in the presence of multiple relays and an eavesdropper. In~\cite{Fan-2017}, impact of the correlated channel in the presence of multiple AF relays and an eavesdropper was considered where lower bound of secrecy outage probability was derived. It was shown that the outage probability improves when full relay selection is carried out as compared to partial relay selection. Furthermore, in~\cite{Xu-2020}, the source sends data to the end device in the presence of multiple eavesdroppers on a slowly fading channel. It was shown that the secrecy rate was improved when the receiver side is highly correlated. All these above-mentioned papers~\cite{Park-2013, Zou-2013,Fan-2017,Xu-2020}, consider that the devices are located midway between source and end device, and equal power allocation is assumed. However, none of these techniques employed distance constraints and applied optimal power allocation. All the techniques mentioned above assume equal power allocation and equal distance between the source and end devices.
\subsection{Contributions}
In this paper, we have proposed a double threshold-based ODS scheme with the input and output threshold. The input threshold helps to select the number of optimal devices depending upon the instantaneous SNR between the source and selected device. The output threshold is applied at the combiner of the end device and this will assist in avoiding the selected devices that experience high fading between the selected device and end device, achieving higher secrecy capacity. Furthermore, as discussed above, all the existing ODS-based schemes assumed the devices/relays to be at the midway between the source and the end device. As the devices have equal distances between each other, equal power assumption is considered for the system. However, in this work, the distance is categorized into three cases and the power of the source and device nodes are varied based on the different distances to replicate the physical D2D environmental conditions. The BER performance of the proposed double threshold-based ODS scheme is analyzed with different modulation schemes and different number of devices. Finally, the intercept probability performance of the proposed scheme is presented in this work. 

The remainder of this paper is organized as follows. The system model of the proposed double threshold-based optimal relay scheme is presented in Section~\ref{sec:System-Model} and the distance-based performance analysis is presented in Section~\ref{sec:distance}. The secrecy capacity and intercept probability expressions for optimal device selection are presented in Section \ref{sec:Secrecy Capacity and Intercept Probability}. Simulation results are presented in Section~\ref{sec:Simulations} and Section~\ref{Sec:Conclusion} concludes the paper.  
\section{SYSTEM MODEL}~\label{sec:System-Model}
\begin{figure}[!t]
\centering
\includegraphics[width=1.0\linewidth]{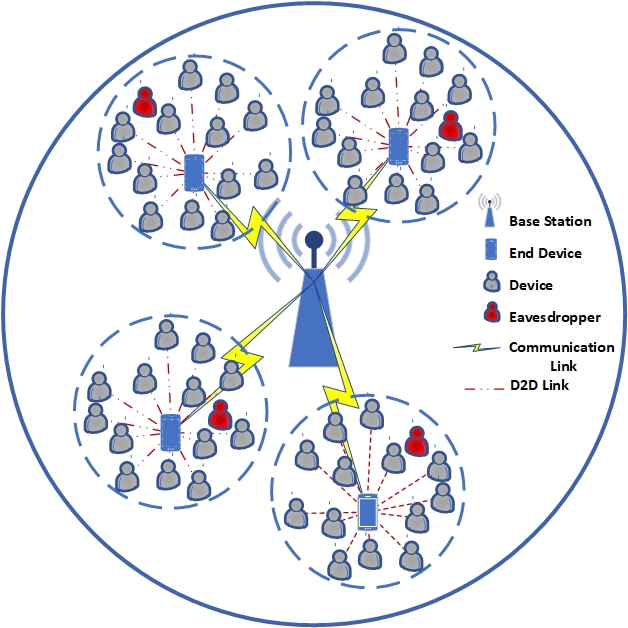}
\caption{Cooperative communication system}
\label{fig:fig-1}
\end{figure}

In Fig.~\ref{fig:fig-1}, base station $S$ transmits the information to the end device, $ED$. There are a number of devices, denoted as $D_i$, that assist in the transmission of the information. Generally, these devices are represented as $D_1, D_2, \cdots, D_N$. In order to design and develop robust threshold mechanism,  we have considered the presence of an eavesdropper, $E$. Passive eavesdropper is considered in this paper, where $E$ is located near to the devices node which is involved in accessing the source information. The assumptions of the proposed model are as follows; all  devices use single antenna each and half-duplex, downlink transmission is considered. The communications takes place during two phases. In Phase-$I$, source $S$ broadcasts the transmitted signal $x$, that is received by the devices  $D_i$, $E$ and $ED$. The set of $N$ devices are selected based on input threshold $\gamma_{i}$. In Phase-$II$ the amplified signal is forwarded by the set of $N$ selected devices to the $ED$ . Out of these $N$ selected devices, some will be dropped depending upon the output threshold $\gamma_o$. Both these thresholds will be discussed in detail in the following section. 
\subsection{Phase-$I$}
The signal received at $ED$ through the direct link in Phase-$I$ is given as~\cite{Abuzaid-2014, Abuzaid-2015,Ahmed-2014, Ahmed-2013}
\begin{equation}~\label{eq-1}
    y_{S,ED} = \sqrt{P_s}h_{S,ED}~x + n_{S,ED}, 
\end{equation}
where $P_s$ denotes the average source power of the transmitted signal. 

The signal received at the device, $D_i$, in Phase-$I$ is given as
\begin{equation}~\label{eq-2}
    y_{S,D_i} = \sqrt{P_s}h_{S,D_i}~x + n_{S,D_i}, \quad i = 1,2,\cdots,N, 
\end{equation}
In equation (\ref{eq-1}) and (\ref{eq-2}), $n_{S,ED}$ and $n_{S,D_i}$ are complex additive white Gaussian noises with zero-mean and variance $N_0$. Similarly, the channel between source and the end device $h_{S,ED}$ and source to the $i$-th device node $h_{S,D_i}$ is assumed to be Rayleigh faded.
\begin{itemize}
\item {\bf Input Threshold, $\gamma_i$:} The input threshold $\gamma_i$ depends on the distance between $S$ and the $i$-th device. The instantaneous SNR of the source to the $i$-th device is given as
\begin{equation}~\label{eq-3}
    \gamma_{S,D_i}=\frac{P_s\zeta_{S,D_i}^2}{N_0}, \quad i=1,2,\cdots,N,
\end{equation}
where $\zeta_{S,D_i}^2=|h_{S, D_i}|^2$ is the channel gain between the source and the $i$-th device. The $i$-th device compares the instantaneous SNR as calculated in (\ref{eq-3}) with this input threshold $\gamma_i$. If the instantaneous SNR is above $\gamma_i$, then the device is selected. If the instantaneous SNR of the device is lower than the input threshold the device is not selected. The selection condition of the $i$-th device can be represented as
\begin{eqnarray}~\label{eq-4}
\textrm{Device, $D_i$}=\left\{
\begin{matrix}
\gamma_{S, D_i}\geq \gamma_{i} &\textrm{Selected}, \\
\gamma_{S, D_i}< \gamma_{i} & \textrm{Not Selected}.
\end{matrix}
\right.
\end{eqnarray}
\end{itemize}
\subsection{Phase-$II$}
\begin{figure*}[!tbh]
\begin{eqnarray}~\label{eq-9}
\setcounter{equation}{9}
\textrm{Selected Transmission Link}=\left\{ 
\begin{matrix}
& \gamma_{S,ED}+\gamma_{D_i, ED}, &\gamma_{S, ED} >\gamma_{o}  ~~\textrm{and}~~\gamma_{S, D_i}>\gamma_{i}, \\
& \gamma_{S, ED}, & \gamma_{S, D_i}<\gamma_{i}\\
& \gamma_{D_i, ED}, & ~~\gamma_{S, ED}<\gamma_{o},\\
& 0, &\gamma_{S, ED}<\gamma_{o}~~ \textrm{and}~~ \gamma_{D_i,ED}<\gamma_{o}\\
\end{matrix}
\right.
\end{eqnarray}
\end{figure*}
In this phase, the $N$ selected devices employ the AF protocol. The amplification factor at the selected $i$-th device is given as~\cite{Laneman-2004, Ahmed-2013}
\begin{equation}~\label{eq-5}
\setcounter{equation}{5}
\alpha_{i}=\frac{\sqrt{P_{D_i}}}{\sqrt{P_s\zeta_{S,D_i}^2+N_0}},\quad i=1,2,\cdots,N,
\end{equation}
where $P_{D_i}$ is the average power applied by the $i$-th device $D_i$. The received signal at the end device is given as
\begin{equation}~\label{eq-6}
y_{D_i, ED} = \alpha_i h_{D_i, ED}~y_{s, D_i}+n_{D_i, ED},
\end{equation}
where $n_{D_i, ED}$ is the complex additive white Gaussian noise with zero-mean and variance $N_0$ and $h_{D_i,ED}$ is the channel coefficient between the $i$-th device and the end device. At the end device, we consider two different types of ODS schemes. In the first type, the wiretap link between the $i$-th device and the eavesdropper $E$ is considered while in the second type, the wiretap link is ignored. 
\begin{itemize}
\item {\bf ODS in the presence of Eavesdropper:} The ODS based scheme in the presence of Eavesdropper (ODS-P) that considers the wiretap channel, $S$, $E$, $D_i$, and $ED$ is given as, 
\begin{eqnarray}~\label{eq-7}
    \textrm{ODS-P} = arg \max_{i \in D_N} \frac{1+\frac{P_S \zeta_{S,D_i}^2\zeta_{D_i,ED}^2}{2(\zeta_{S,D_i}^2+\zeta_{D_i,ED}^2)N_0}}{1+\frac{P_S \zeta_{S,D_i}^2\zeta_{D_i,E}^2}{2(\zeta_{S,D_i}^2+\zeta_{D_i,E}^2)N_0}},
\end{eqnarray}
where $\zeta_{D_i,ED}^2=|h_{D_i, ED}|^2$ and $\zeta_{D_i,E}^2=|h_{D_i, E}|^2$ are the channel gains of the $i$-th device to the $ED$ and to the eavesdropper $E$, respectively. 

\item{\bf ODS in the absence of Eavesdropper:} In the ODS based scheme that does not consider the wiretap link (i.e., the absence of eavesdropper), (ODS-A), maximum capacity from the $i$-th device to end device, $ED$, is given as ~\cite{Zou-2013}
\begin{eqnarray}~\label{eq-8}
    \textrm{ODS-A} = arg \max_{i \in D_N} \frac{\zeta_{S,D_i}^2\zeta_{D_i, ED}^2}{\zeta_{S,D_i}^2+\zeta_{D_i,ED}^2}.
\end{eqnarray}
\item {\bf Output Threshold:} The signal at the end device, $ED$, will be weak and the output signal will be degraded if the link between the device $D$ and the end device, $ED$, has high fading. The output threshold at the combiner  selects the best device to end device link from the $i$-th device to the end device. The threshold set at the output of the combiner $\gamma_o$ tests the effect of selected $i$-th device and the signal from the direct link. Output threshold testing condition is given in (\ref{eq-10}).
\end{itemize}
\section{Distance Analysis}~\label{sec:distance}
The proposed device selection scheme is analysed based on the distance calculated between the source $S$, selected device $D_i$, and end device $ED$. The total power $P_t$ of the system is defined as 
\begin{equation}~\label{eq-10}
\setcounter{equation}{10}
P_t=P_S+\sum_{i=1}^N P_{D_i}. 
\end{equation}
%
\subsection{Case-I}~\label{sec:Case-I}
In this case the devices are assumed to be in the midway between the source $S$ and end device $ED$. Furthermore, the eavesdropper $E$ is assumed to be located near to the selected $N$ devices. The distance expression can be represented as
\begin{equation}~\label{eq-11}
    d(S,D_i)=d(D_i, ED).
\end{equation}
As the source power is fixed to $P_S=0.5$, all the selected devices $D_i$ are given the half of the remaining  power.  
\subsection{Case-II}~\label{sec:Case-II}
In this case, the devices are assumed to be away from $S$ and nearer to $ED$. The eavesdropper $E$ is assumed to be away from the $S$ and near to the selected devices. The distances can be represented as
\begin{equation}~\label{eq-12}
    d(S,D_i)>>d(D_i, ED).
\end{equation}
The power ratios are calculated using the optimum power allocation for AF based cooperative systems at high SNR in Theorem 5.3.4 ~\cite{ Liu-2010}
\begin{eqnarray}~\label{eq-13,14}
P_S &=& \frac{\sigma_{S,D_i}+\sqrt{\sigma_{S,D_i}^2+8\sigma_{D_i,ED}^2}}{3\sigma_{S,D_i}+\sqrt{\sigma_{S,D_i}^2+8\sigma_{D_i,ED}^2}}P_t,\\
P_{D_i} &=& \frac{1}{N}\cdot\frac{2\sigma_{S,D_i}}{3\sigma_{S,D_i}+\sqrt{\sigma_{S,D_i}^2+8\sigma_{D_i,ED}^2}}P_t.
\end{eqnarray}
The channel link quality between $D_i$ and $ED$ will be better as compared to the $S, D_i$ link due to lesser distance, the power ratios is calculated by using $\sigma_{S,D_i}^2=1$ and $\sigma_{D_i, ED}^2=10$ where $\sigma_{S,D_i}^2$ is the variance of channel link from the source to $i$-th device and $\sigma_{D_i, ED}^2$ is the channel link from the $i$-th device to the end device. 
\subsection{Case-III}~\label{sec:Case-III}
The devices are assumed to be nearer the $S$ and away from the $ED$. The eavesdropper $E$ is assumed to be near to the $S$. The distance expression is given as
\begin{equation}~\label{eq-15}
    d(S,D_i)<<d(D_i, ED)
\end{equation}
The power ratios are calculated as 
\begin{eqnarray}~\label{eq-16,17}
P_S &=& \frac{2\sigma_{S,D_i}}{3\sigma_{S,D_i}+\sqrt{\sigma_{S,D_i}^2+8\sigma_{D_i,ED}^2}}P_t,\\
P_{D_i}&=&\frac{1}{N} \cdot \frac{\sigma_{S,D_i}+\sqrt{\sigma_{S,D_i}^2+8\sigma_{D_i,ED}^2}}{3\sigma_{S,D_i}+\sqrt{\sigma_{S,D_i}^2+8\sigma_{D_i,ED}^2}}P_t,
\end{eqnarray}
As mentioned above, the power ratios are calculated by using $\sigma_{S,D_i}^2=10$ and $\sigma_{D_i, ED}^2=1$.

\section{Secrecy Capacity and Intercept Probability}~\label{sec:Secrecy Capacity and Intercept Probability}
The secrecy capacity of the direct path is calculated as the difference between the capacity of the source to destination and the wiretap link [7] and is given as
\begin{equation}~\label{eq-18}
SEC_{direct} = C_{S,D}^{direct} - C_{S,E}^{direct}.
\end{equation}

The capacity of direct path from S to ED and S to E is given as 
\begin{equation}~\label{eq-19}
C_{S,ED}^{direct} = \log_2 \left({1+\frac{\zeta_{S,ED}^2 P_t}{{N_0}}}\right), 
\end{equation}
\begin{equation}~\label{eq-20}
C_{S,E}^{direct} = \log_2 \left({1+\frac{\zeta_{S,E}^2 P_t}{N_0}}\right).
\end{equation}
where $\zeta_{S,ED}^2 = |h_{S,ED}|^2$ and $\zeta_{S,E}^2 = |h_{S,E}|^2$.

The secrecy capacity will be high, if the capacity of the direct link is more than the wiretap link. The eavesdropper cannot intercept the source signal in this case. The intercept probability is defined as the probability of successfully interception of the source signal by the eavesdropper. This is possible if the secrecy capacity of the link is less than wiretap link. It is considered an important performance metric of physical-layer security. 
The intercept probability of direct path [13] is given as
\begin{equation}~\label{eq-21}
IP_{direct} = \frac {\rho_{S,E}^2}{{\rho_{S,E}^2}+{{\rho_{S,ED}^2}}},
\end{equation}
where $\rho_{S,E}^2 = E{(|h_{S,E}|^2)}$ and $\rho_{S,ED}^2 = E{(|h_{S,ED}|^2)}$.

The intercept probability expression of ODS [7] is given as 
\begin{equation}~\label{eq-22}
IP_{ODS} = \prod_{i=1}^M \frac{\rho_{D_i,E}^2}{({\rho_{D_i,E}^2} + {\rho_{D_i,ED}^2})}. 
\end{equation}
where $\rho_{D_i,E}^2 = E(|h_{D_i,E}|^2)$ and $\rho_{D_i,ED}^2 = E(|h_{D_i,ED}|^2)$.
\section{Simulation Results}~\label{sec:Simulations}
In this section, the performance improvement of the proposed double threshold ODS scheme is validated by simulations. The Monte-carlo simulations are performed and the number of bits considered for the simulation is equal to $10^5$ while the number of samples per symbol used is $10$ for convenience~\cite{Daghal-2016}. Passive eavesdropper $E$ is considered in these simulations where we consider the presence (PODS-P) and absence (PODS-A) of eavesdropper in our simulation. The PODS-P is achieved by combining the ODS scheme with input threshold at the device,$D_i$, and the output threshold at the End Device, $ED$, with the optimum power allocation for the three different cases in the presence of the passive eavesdropper near to the device, $D_i$. The PODS-A scheme is achieved in the similar way without the wiretap link. We compare our scheme with direct path, ODS scheme without threshold as proposed in~\cite{Zou-2013}. For the direct path simulations only path between $S$ and $ED$ is chosen. No cooperation is assumed between the Source $S$, $N$ intermediate devices, and end device, $ED$. The simulation results are based on the different distance cases, with various device scenarios, and modulation schemes. We first study the effect of distance on the proposed scheme. 
\subsection{Performance based on Different Distances}
\begin{figure}[!t]
    \centering
    \includegraphics[width=1.0\linewidth]{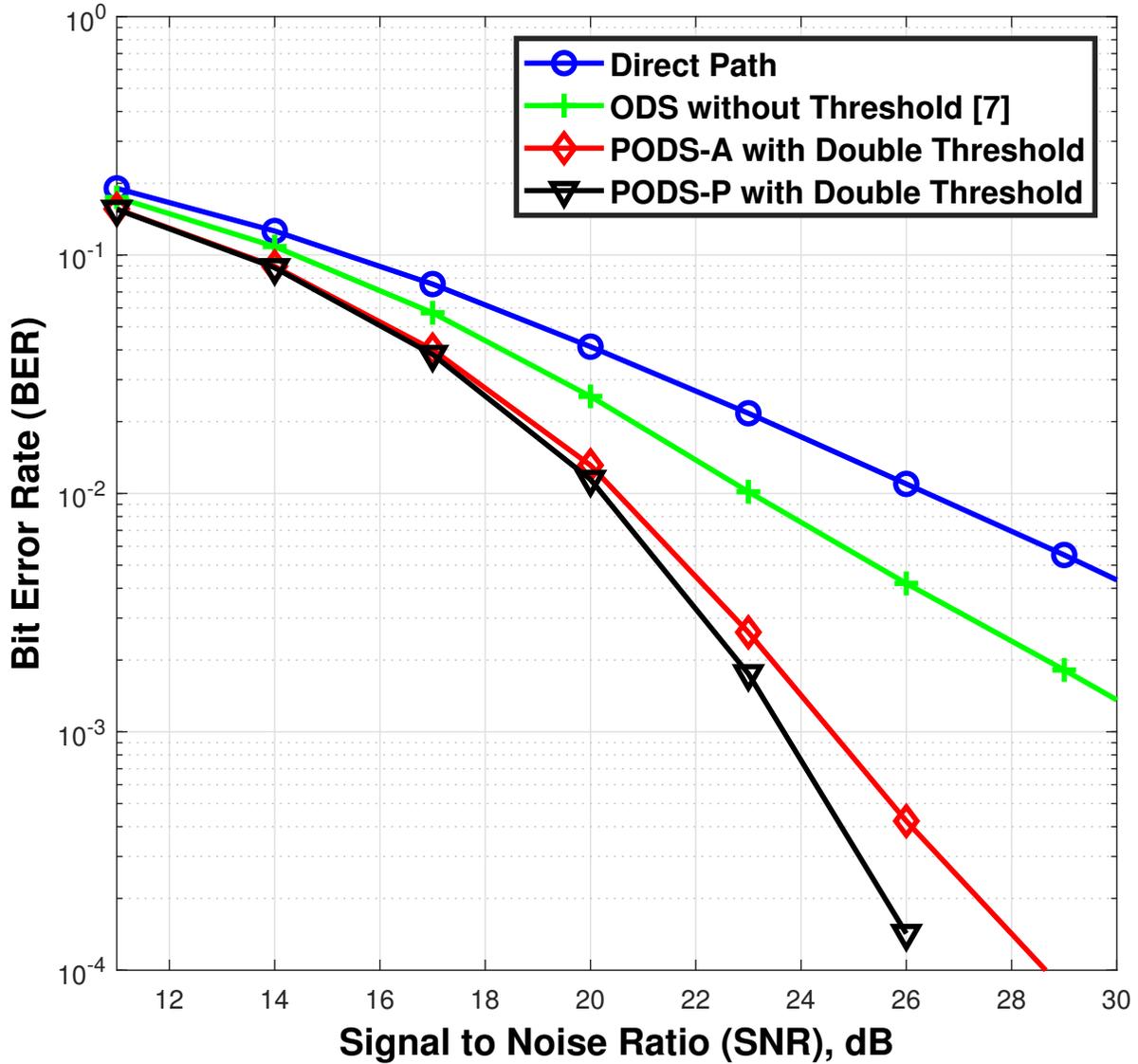}
    \caption{BER performance comparison of the proposed device selection scheme for Case-I}
    \label{fig:fig-2}
\end{figure}

In Fig.~\ref{fig:fig-2}, the bit error rate (BER) versus signal-to-noise ratio (SNR) is plotted for the Case-I as discussed in Section \ref{sec:Case-I}.  
In the ODS-scheme without threshold, all the devices are selected based on optimal relay selection criteria as in ~\cite{Zou-2013} and the direct path between the source $S$ and end device $ED$ does not exist. It can be observed that there is a significant gain as compared to direct path, especially at higher SNRs. A gain of more than $4$~dB is achieved at a BER of $0.004$ when comparing the performance between the direct path communication and ODS without threshold. In the PODS-A and PODS-P scenarios, the input threshold and  output threshold are $\gamma_i = 5$~dB and  $\gamma_o= 5$~dB, respectively. In the third scenario, the PODS-A scheme is employed with a double threshold and no direct path exists between the $S$ and $ED$. While in the fourth scenario, the proposed ODS (PODS-P)-scheme is selected with a double threshold and there is a direct path between source and destination.  The difference of $2$~dB is observed between the PODS-P and PODS-A with double threshold at the BER of approximately $10^{-4}$. It is seen that the performance of the PODS-P and PODS-A significantly outperforms the ODS-scheme without threshold and the direct path scheme. 

\begin{figure}[!t]
    \centering
    \includegraphics[width=1.0\linewidth]{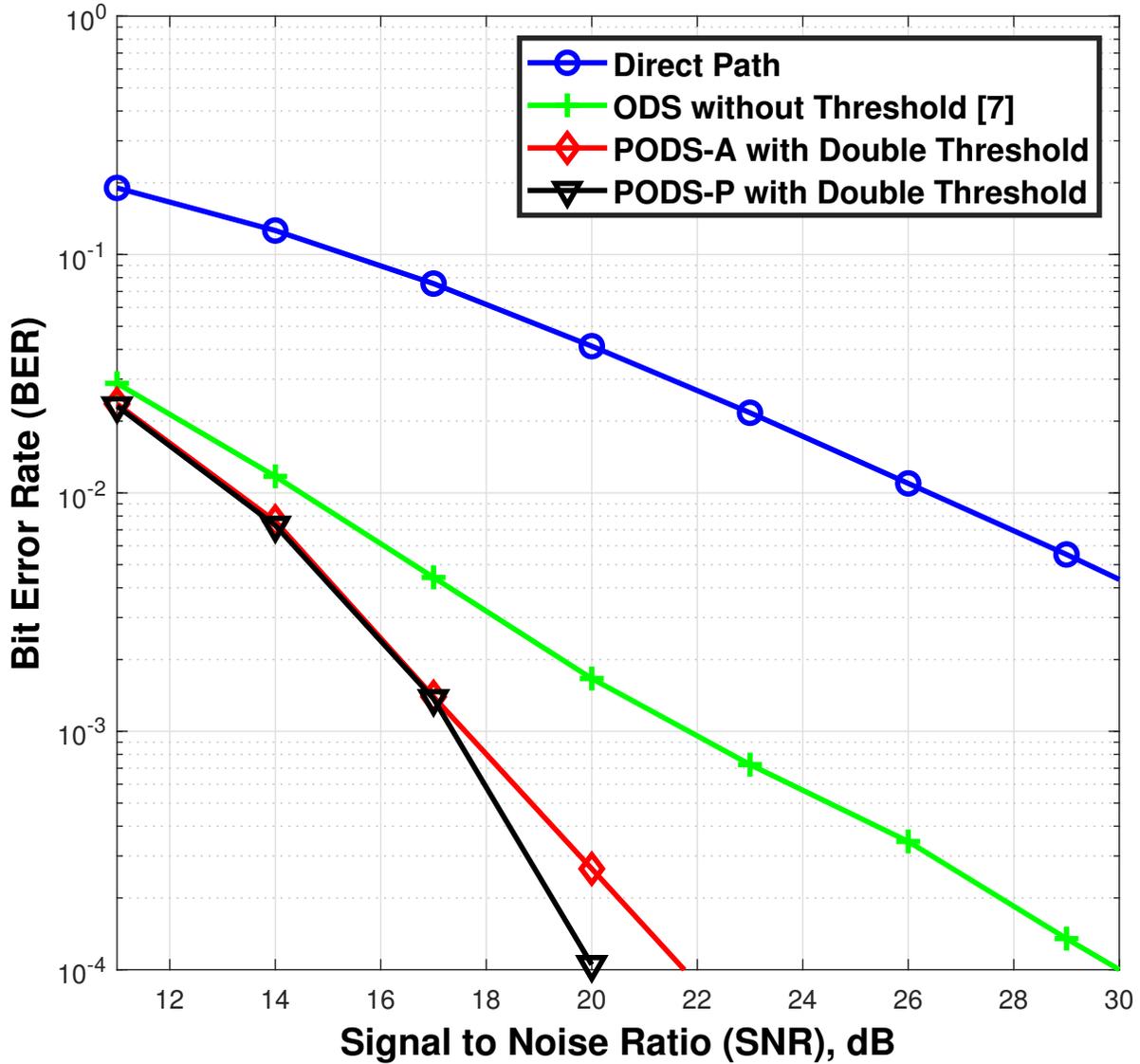}
    \caption{BER performance comparison of the proposed device selection scheme for Case-II}
    \label{fig:fig-3}
\end{figure}

Fig.~\ref{fig:fig-3} shows the performance of the PODS-A and PODS-P where double threshold is compared with the existing ODS without threshold and direct path for Case-II as described in Section~\ref{sec:Case-II}. The input threshold $\gamma_i$ at the devices will be less as the devices are away from the $S$. However, a higher output threshold value $\gamma_o$ at the $ED$ will be selected as the devices are near to $ED$. The input threshold and the output threshold are $\gamma_i=5$~dB and $\gamma_o = 10$~dB, respectively. From the simulation results a significant gain of more than $6$~dB in BER is achieved as compared to Fig.~\ref{fig:fig-2} due to higher output threshold for both the proposed PODS-A and PODS-P schemes. This gain in performance is due to closer distance between the devices and the $ED$. A significant performance improvement is observed in ODS without threshold scheme as compared to that of the direct path. However, the performance of our schemes are much superior than proposed scheme in~\cite{Zou-2013} and the direct path scheme.

\begin{figure}[!t]
    \centering
    \includegraphics[width=1.0\linewidth]{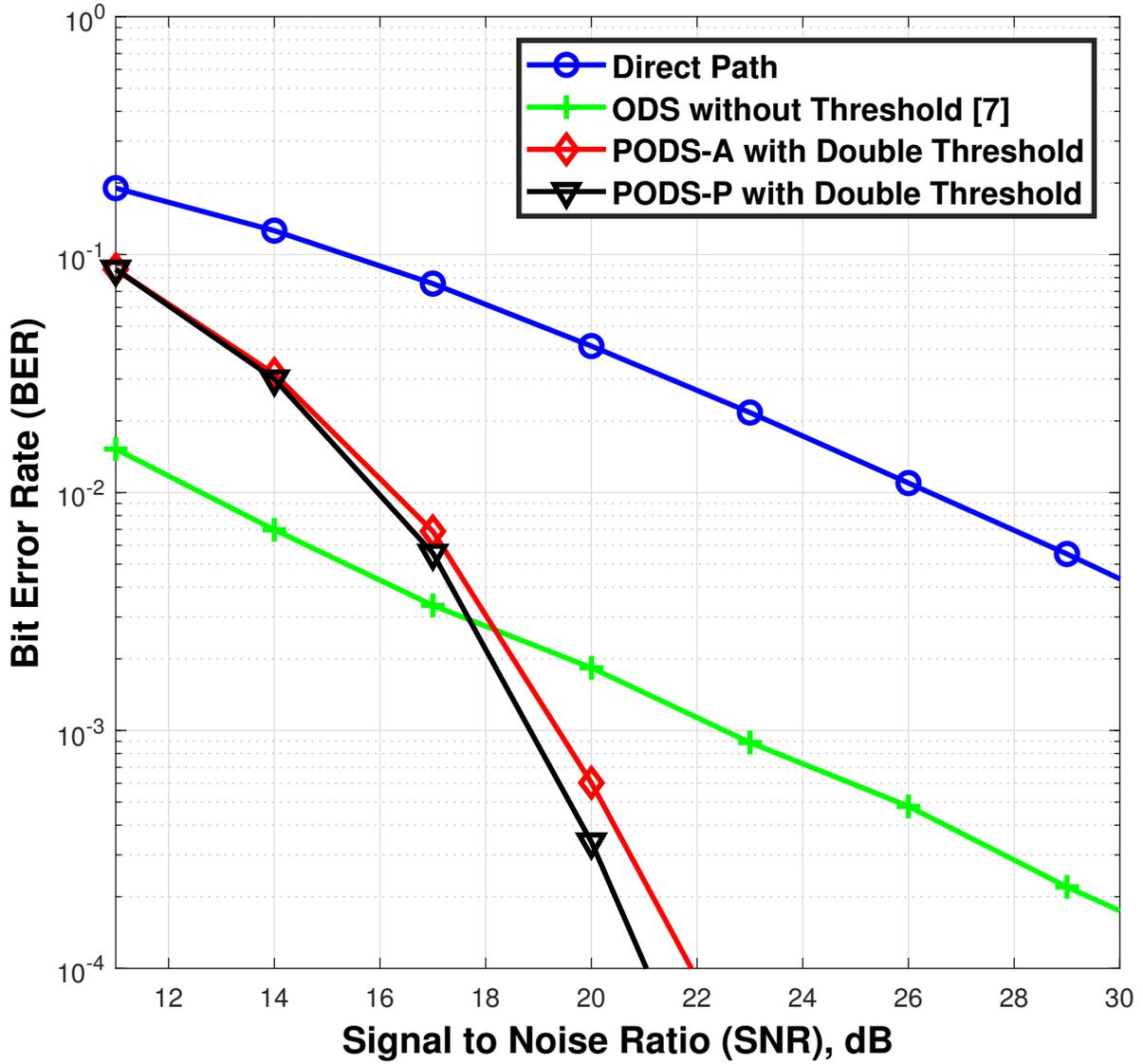}
    \caption{BER performance comparison of the proposed device selection scheme for Case-III}
    \label{fig:fig-4}
\end{figure}

Fig.~\ref{fig:fig-4} shows the performance comparison of the PODS-A and PODS-P schemes with double threshold when compared with the existing ODS without a threshold and direct path for Case-III as described in Section~\ref{sec:Case-III}. The proposed schemes outperform the direct path scheme. However, at lower SNR values the ODS scheme without threshold outperforms our proposed PODS-A and PODS-P schemes. In this scenario, the ODS without threshold scheme employs all the nearby devices that receive the broadcast information from the $S$ and as a result all the devices will forward the information to $ED$, therefore, a significant gain is observed at the low SNR. Our proposed PODS-A and PODS-P perform poorly as compared to ODS without threshold because of high input threshold value as in this case devices are very close to the $S$ as described in Section~\ref{sec:Case-III}. It is seen that at SNRs more than $20$~dB, our proposed scheme outperforms the other schemes.

\subsection{Performance comparison based on Different Modulation}
\begin{figure}[!t]
    \centering
    \includegraphics[width=1.0\linewidth]{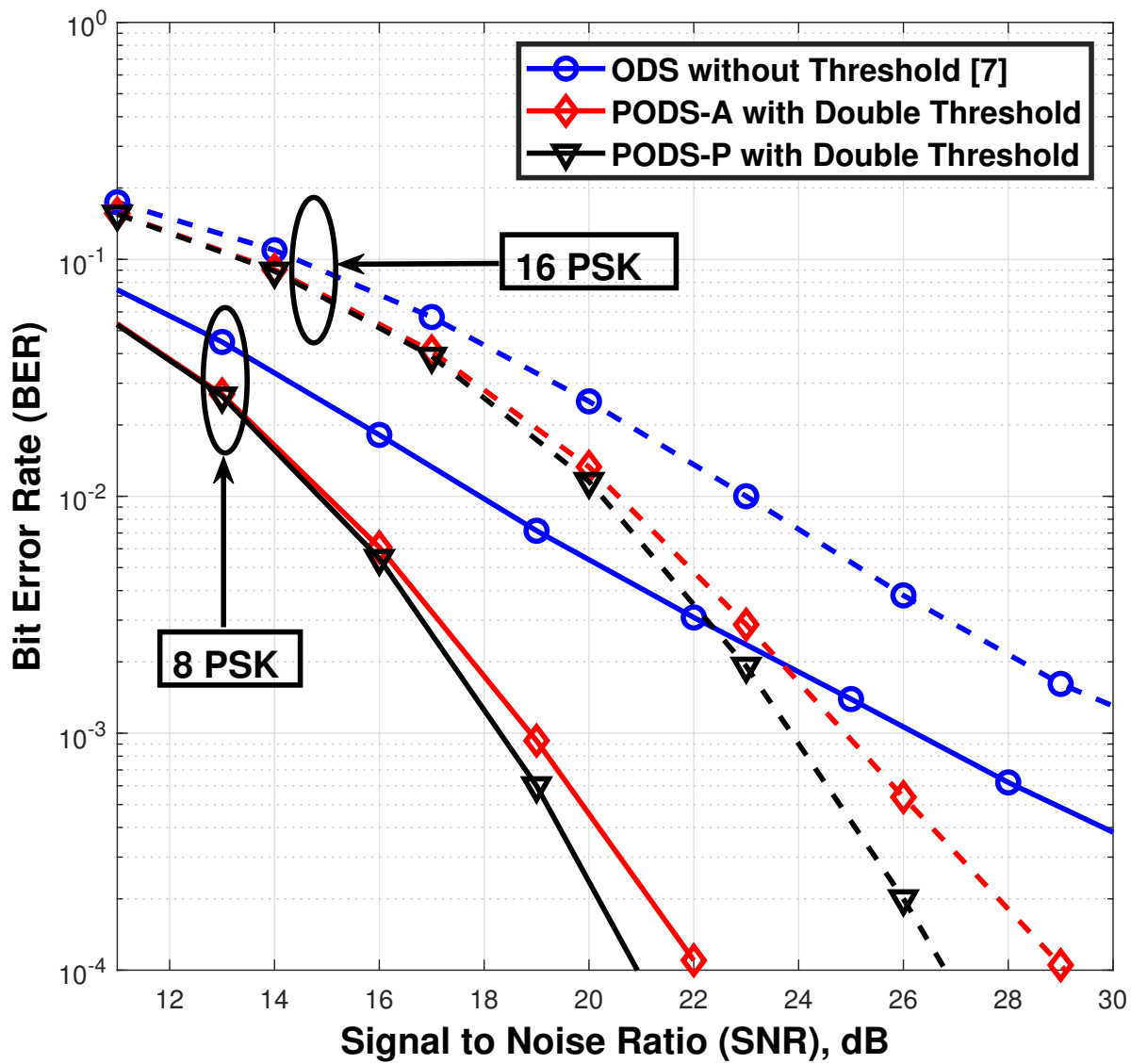}
    \caption{Comparison of the proposed device selection scheme with $8$-PSK and $16$-PSK for Case-I}
    \label{fig:fig-5}
\end{figure}

In Fig.~\ref{fig:fig-5}, performance comparison of the proposed PODS-A and PODS-P with double threshold is compared with the existing ODS without threshold scheme for $8$-PSK and $16$-PSK modulation schemes. The BER performance curve shows an improvement for the $8$-PSK modulation compared to $16$-PSK modulation for all the considered schemes. The BER performance of $8$-PSK modulation achieves an improvement of more than $6$~dB at a BER of $0.003$ compared to $16$-PSK modulation. A gain of more than $7$~dB is achieved at a BER of $0.003$ for PODS-A with double threshold and PODS-P with double threshold compared to the ODS without threshold. Thus, it can be observed that our scheme is adaptable and will be applicable to higher PSK modulation.

\subsection{Performance comparison based on Different Number of Devices}
In Fig. 6, the performance of the proposed schemes is compared to increased number of devices in the system. In all our previous simulations, the number of devices were fixed to $5$.  However, here, we have increased the number of devices to $10$. Three different curves are considered: PODS-A with double threshold, PODS-P with double threshold and ODS without threshold. It can be observed that there is a significant gain in BER performance as the number of devices are increased for both the proposed PODS-A and PODS-P with double threshold. However, for ODS without threshold no significant gain is observed. From this simulation, it can be observed that our system can scale easily with the increased number of devices. However, the ODS scheme without threshold cannot scale with the number of devices.

\begin{figure}[!t]
    \centering
    \includegraphics[width=1.0\linewidth]{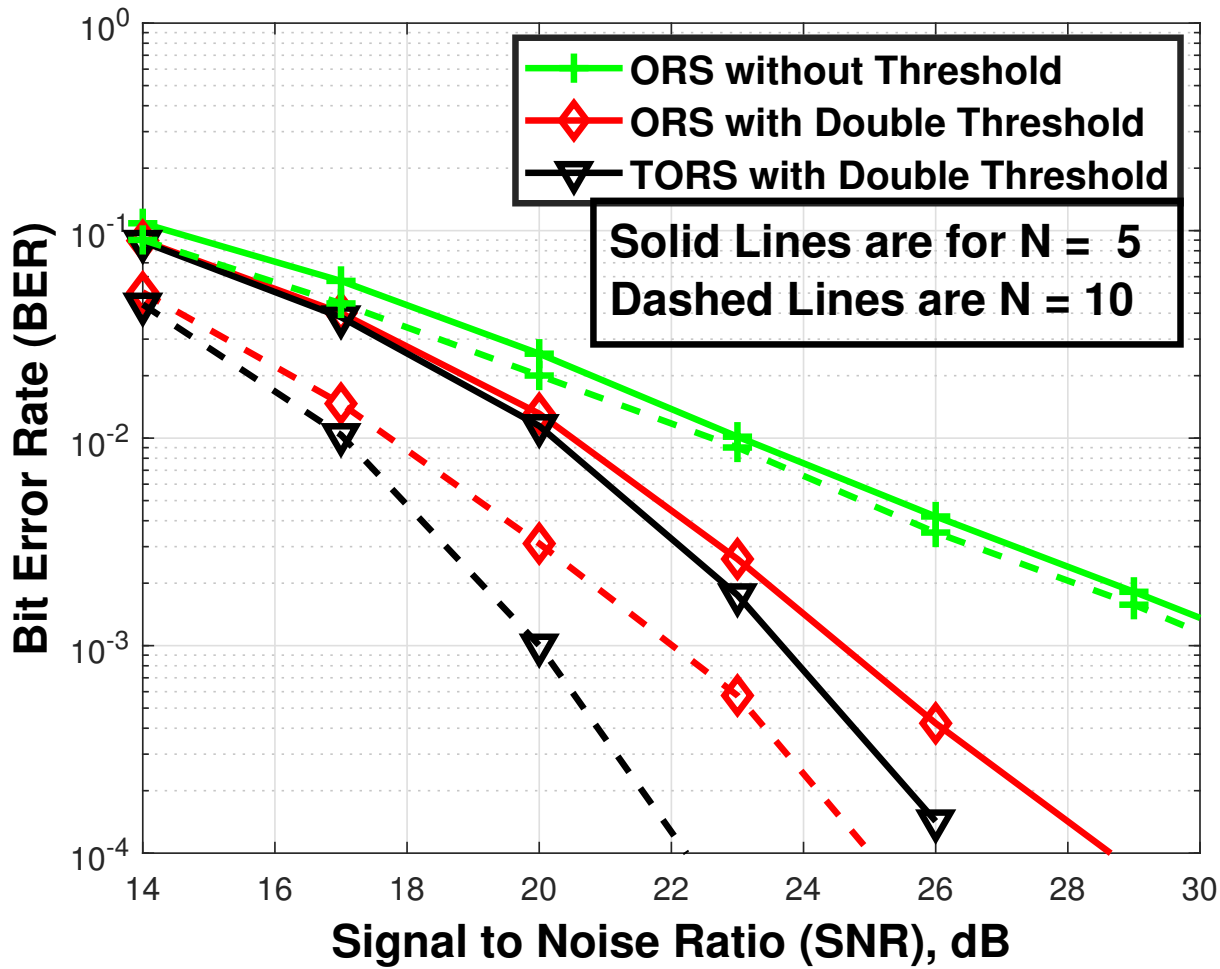}
   \caption{Comparison of the proposed device selection scheme with $5$ and $10$-relay for Case-I}
    \label{fig:fig-6}
\end{figure}

\subsection{Intercept probability Performance }
In Fig. 7, the intercept probability performance of the proposed PODS-A and PODS-P with double threshold is compared with direct path. The number of devices was fixed to $5$. Three different curves are considered: PODS-A with double threshold, PODS-P with double threshold and direct path. It can be observed that there is a significant performance improvement observed for the PODS-P with double threshold compared to PODS-A with double threshold and direct path. This improvement in intercept probability performance confirms the physical layer security improvement of the proposed scheme. Intercept probability performance of the PODS-P with double threshold is high due to the increase in the secrecy capacity and this ensures the probability of the eavesdropper intercepting the source information being smaller and thereby the proposed ODS scheme enhances the physical layer security.

\begin{figure}[!t]
    \centering
    \includegraphics[width=1.0\linewidth]{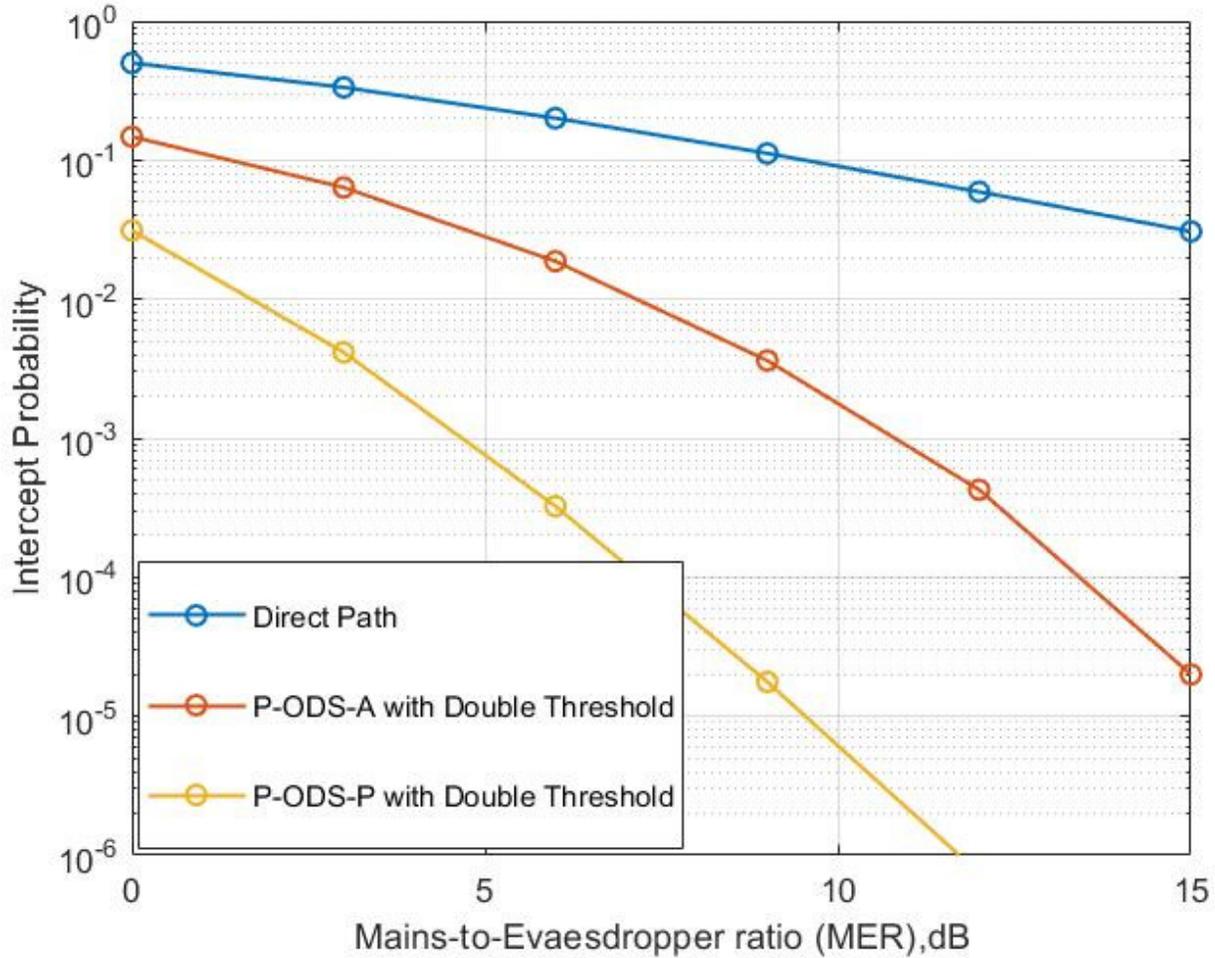}
     \caption{Intercept probability performance of the proposed device selection scheme with $5$-relay for Case-I}
    \label{fig:fig-7}
\end{figure}

\section{Conclusion}~\label{Sec:Conclusion}
In this paper, the BER performance of the proposed double threshold-based optimal device selection scheme was analyzed, where the distance between the source, devices, and end device was taken into consideration. The effect of presence and absence of direct path was also considered. The comparison of the proposed scheme with the existing optimal device selection scheme for the different cases shows considerable performance improvement. The BER performance of the proposed scheme shows improvement for both $8$-PSK and $16$-PSK schemes compared to the existing scheme. For different device scenarios, the BER performance improvement is observed for the proposed schemes as compared to the existing scheme without threshold which did not scale with the higher number of devices. The intercept probability results shows significant improvement in the performance of the  proposed scheme and ensures physical layer security improvement.

\vspace{12pt}

\end{document}